# Three-dimensional quantum Griffiths singularity in bulk iron-pnictide superconductors

Shao-Bo Liu[1†], Congkuan Tian[1,2†], Yongqing Cai[3,4†], Hang Cui[1], Xinjian Wei[2], Mantang Chen[1], Yang Zhao[1], Yuan Sui[1], Shuyue Guan[1], Shuang Jia[1], Yu Zhang[2], Ya Feng[2], Jiankun Li[2], Jian Cui[2], Yuanjun Song[2], Tingting Hao[2], Chaoyu Chen[4], Jian-Hao Chen [1,2,5,6*]

[1] International Center for Quantum Materials, School of Physics, Peking University, Beijing 100091, China
[2] Beijing Academy of Quantum Information Sciences, Beijing 100094, China
[3] School of Physics, Dalian University of Technology, Dalian 116024, China
[4] Shenzhen Institute for Quantum Science and Engineering and Department of Physics, Southern University of Science and Technology, Shenzhen 518055, China.
[5] Key Laboratory for the Physics and Chemistry of Nanodevices, Peking University, Beijing 100091, China
[6] Hefei National Laboratory, Hefei 230026, China

[†]These authors contributed equally to this work.
Corresponding Author:
*E-mail: Jian-Hao Chen (chenjianhao@pku.edu.cn)

## Abstract

**The quantum Griffiths singularity (QGS) is a phenomenon driven by quenched disorders that break conventional scaling invariance and result in a divergent dynamical critical exponent during quantum phase transitions (QPT). While this phenomenon has been well-documented in low-dimensional conventional superconductors and in three-dimensional (3D) magnetic metal systems, its presence in 3D superconducting systems and in unconventional high-temperature superconductors (high-$T_c$ SCs) remains unclear. In this study, we report the observation of robust QGS in the superconductor-metal transition (SMT) of both quasi-2D and 3D anisotropic unconventional high-$T_c$ superconductor $CaFe_{1-x}Ni_xAsF$ ($x < 5\%$) bulk single crystals, where the QGS states persist to up to 5.3 K. A comprehensive quantum phase diagram is established that delineates the 3D anisotropic QGS of SMT induced by perpendicular and parallel magnetic field. Our findings reveal the universality of QGS in 3D superconducting systems and unconventional high-$T_c$ SCs, thereby substantially expanding the range of applicability of QGS.**

**Introduction**

The superconductor-insulator (metal) transition (SIT/SMT), a prototypical example of quantum phase transition (QPT), has garnered significant attention due to its implications for understanding quantum states of matter as well as its potential applications in novel low-dimensional superconducting quantum computing devices[1-7]. In conventional SIT/SMT systems, a single quantum critical point (QCP) with power-law divergence of a single spatial or temporal correlation length as a function of non-thermal control parameters is presented[2,8,9]. The critical exponents of the divergence reflect the universality class of the quantum critical behavior[10,11]. However, the presence of quenched disorders can disrupt conventional scaling invariance and alter the universality class of the QPT, leading to the emergence of the quantum Griffiths phase, which is characterized by continuously varying critical exponents with respect to temperature and control parameters, fundamentally challenging the conventional understanding of QPT in homogeneous disordered systems[12-16]. In the quantum Griffiths phase, distinct superconducting islands ("rare regions") emerge and the fluctuations of the order parameter within these rare regions become non-negligible[12-16]. QGS have been experimentally observed in two-dimensional (2D)[5,17-23] and quasi-one-dimensional (quasi-1D)[7] conventional superconductors. However, the fate of QGS in three-dimensional bulk superconductors and unconventional high-$T_c$ superconductors remains unknown.

The 1111-type FeAs-based superconductors are the earliest discovered iron-based unconventional high-$T_c$ superconductors with the highest $T_c$ in the material family[24]. CaFeAsF, a representative parent compound of the 1111-type FeAs-based superconductors, is an antiferromagnetic bad-metal consisting of alternately stacked CaF and FeAs layers along the $c$-axis[24]. Quantum oscillation measurements in CaFeAsF, along with band-structure calculations, revealed a pair of symmetry-related Dirac electron cylinders and a normal hole cylinder at the Fermi surface[25]. Nontrivial topological electronic structure has been predicted in CaFeAsF arising from strong electronic correlations of the Fe 3d electrons[26] and superconductivity in CaFeAsF can be achieved through chemical doping or external pressure[24,27,28]. Although undoped CaFeAsF is not superconducting at ambient pressure, a magnetic-field-induced metal-insulator QPT near the quantum limit has been reported[29], providing a plausible precursor for SIT/SMT in doped and superconducting CaFeAsF. Electron doping of CaFeAsF can be achieved by substituting a fraction of Fe with Ni, resulting in the formation of CaFe$_{1-x}$Ni$_x$AsF ($x \ll 1$). However, single crystal of CaFe$_{1-x}$Ni$_x$AsF has been challenging to grow, impeding further exploration of this material.

In this study, superconducting single crystals of CaFe$_{1-x}$Ni$_x$AsF ($x < 5\%$) is successfully synthesized using the flux method (details in Methods and Supplementary Information S1 and S2). In all the samples where SMTs are realized, multiple QCPs and diverging dynamic critical exponents are observed, providing direct evidence of QGS in these materials[5,12,18,30]. By fitting the experimental divergence behavior of "critical exponent" versus magnetic field to an activated scaling law[5], the value of the extracted exponents are 0.6 and 0.4 for quasi-2D and 3D anisotropic CaFe$_{1-x}$Ni$_x$AsF, respectively, consistent with theoretical predictions and numerical simulations[31-34]. In particular, this is the first experimental observation of $B_\perp$- and $B_{//}$-driven QGS of SMT in a 3D anisotropic superconductor and in an unconventional high-$T_c$ superconductor, which serves to catalyze new research efforts on substantially extended physical grounds of the QGS effect.

**Magnetic-field-driven SMTs with multiple QCPs in CaFe$_{1-x}$Ni$_x$AsF single crystals**

High quality CaFe$_{1-x}$Ni$_x$AsF single crystals were synthesized and characterized as shown in Methods and Supplementary Information S1&S2. For clarity, CaFe$_{1-x}$Ni$_x$AsF samples with $x$ = 3.1%, 3.5%, 3.7% and 4.9% are denoted as sample A, B, C and D, respectively. Figs. 1a-d show the resistance versus temperature $R(T)$ curves of samples A-D under both $B_\perp$ (upper panels) and $B_{//}$ (lower panels) from 0 T to 14 T in log-log scale. The $R(T)$ curves in linear coordinates are shown in Supplementary Information S4. The onset superconducting transition temperature at zero field ($T_c^{onset}$) for sample A, B, C and D are 5.75 K, 6.3 K, 7.5 K and 11 K, respectively. The upper panels of Figs. 1a-d show monotonically decreasing superconducting transition temperature $T_c$ with increasing $B_\perp$, ultimately leading to a weakly insulating $R(T)$ without superconducting transition at arbitrarily low temperatures, i.e., leading to a $B_\perp$-driven SMT. The lower panels of Figs. 1a-d show similar monotonic decrease in $T_c$ with increasing $B_{//}$, but the $B_{//}$-driven SMT is only observed in sample A (Fig. 1a, lower panel). For samples B, C and D (Figs. 1b-d, lower panels), it is evident that $B_{//}$ larger than 14T may be required to fully suppress superconductivity and to achieve SMT. The brown arrows in Figs. 1a-d highlight the $R(T)$ curves at specific "critical" magnetic fields where no onset of superconductivity can be detected at the lowest measured temperature, representing the emergence of magnetic field-driven SMT in each sample. Below these critical magnetic fields, continuously changing crossing points in the $R(B)$ isotherms can be extracted from the $R(T, B)$ data (see Fig. 3), indicative of the presence of multiple QCPs and contrasts with conventional quantum phase transitions with a single QCP[2,8,9].

**Doping tunable quasi-2D to 3D anisotropic crossover in CaFe$_{1-x}$Ni$_x$AsF single crystals**

Electronic dimension is an important character that governs material properties and detectable via transport techniques. Here we show that Ni doping can induce an electronic dimensional crossover in the CaFe$_{1-x}$Ni$_x$AsF single crystals. Among the four samples investigated in this study, sample A has 3D anisotropic electronic structure while samples B, C and D have quasi-2D electronic structures, in both the normal states and the superconducting states. We shall focus on samples A and C in the main text; data from samples B and D can be found in Supplementary Information S5-8. Figs. 2a-b show the normal state magnetotransport data MR vs. $B\cos\theta$ for samples A and C, respectively. Here, the magnetic field ranges from 0 T to 14 T with its angle $\theta$ from 0° (along the $c$-axis) to 90° (along the $a$-axis, defined as the in-plane direction perpendicular to the current) and the temperature $T$ = 14 K > $T_c$. As shown in Fig. 2a, the MR curves of sample A do not scale into a single curve when plotted with $B\cos\theta$, pointing to a three-dimensional Fermi surface in the normal state of sample A. On the contrary, the MR curves of sample C scale remarkably well with $B\cos\theta$ (Fig. 2b), indicating the dominance of two-dimensional Fermi surface in sample C above $T_c$ [35,36]. For the superconducting state, Figs. 2c-d show the angular dependence of the upper critical field ($H_{c2}$) at $T$ = 1.58 K < $T_c$ for sample A and C, respectively. Here, $H_{c2}$ is defined as the magnetic field where the resistance becomes 50% of the normal state resistance, which is extracted from Supplementary Information S13 and S9 for sample A and C, respectively. In Fig. 2c, $H_{c2}(\theta)$ of sample A can be well fitted by the 3D anisotropic mass model[37], where $H_{c2}(\theta) = H_{c2}^{//}/(\sin^2\theta + \gamma^2\cos^2\theta)^{1/2}$ with $\gamma = H_{c2}^{//}/H_{c2}^{\perp}$, but not the 2D Tinkham model[38], where $(H_{c2}(\theta)\sin\theta/H_{c2}^{//})^2 + |H_{c2}(\theta)\cos\theta/H_{c2}^{\perp}| = 1$, indicating that the superconducting state in sample A is still three-dimensional. Here $H_{c2}^{//}$ and $H_{c2}^{\perp}$ represent the in-plane and out-of-plane upper critical field, respectively. In contrast, Fig. 2d shows the $H_{c2}(\theta)$ of sample C that can be fitted by the 2D Tinkham model but not the 3D anisotropic mass model; together with the Berezinskii−Kosterlitz−Thouless (BKT) transition[39] observed in sample C (Supplementary Information S10), it is evident that sample C has a quasi-2D superconducting electronic state. Similar analysis is performed for sample B and D as shown in Supplementary Information S5a and S7a, respectively, also showing quasi-2D electronic structure for the two samples.

**QGS in quasi-2D and 3D anisotropic unconventional high-$T_c$ SCs**

Since QGS has so far been observed in 2D and quasi-1D conventional superconductors, we first investigate the possibility of QGS states in quasi-2D unconventional high-$T_c$ SCs (Sample B, C and D). Fig. 3a shows the MR of sample C versus $B_\perp$ at various temperatures ranging from 0.46 K to 7 K. Indeed, the MR isotherms exhibit a series of continuously moving crossing points ($B_c$, $R_c$) rather than a single point, indicative of QGS behavior. Fig. 3b plots the evolution of the line of "critical" points for the SMTs in terms of $B_c$ versus $T$. Notably, as $T$ decreases, $B_c$ exhibits continuous upward displacement, deviating from the mean field Werthamer-Helfand-Hohenberg (WHH) theory[40], in agreement with previous reports in 2D QGS of SMTs in conventional superconductors[17,18].

To gain a quantitative understanding of the QGS in CaFe$_{1-x}$Ni$_x$AsF samples, we employed finite-size scaling (FSS) analysis to investigate the critical points of the SMTs. The resistance of the sample near these critical points follows a scaling form[1]: $R(\delta, t) = R_c \cdot f[\delta \cdot t(T)]$. Here, $t \equiv (T/T_0)^{-1/zv}$ with $z$ the dynamical critical exponent, $v$ the correlation length exponent and $T_0$ the lowest temperature in the fitting range; $R_c$ and $B_c$ are the critical resistance and critical magnetic field obtained from the critical points, respectively; $\delta = |B - B_c|$ is the deviation from $B_c$ and $f(x)$ is an arbitrary function with $f(0) = 1$. For the quasi-2D sample C, the "critical" point ($B_c$, $R_c$) of a certain small critical transition region is defined as the crossing point of several $R(B)$ curves with adjacent temperatures (details in Supplementary Information S11). The scaling results are presented in Supplementary Information S12 and Fig. 3c, which give the effective "critical" exponents $z\nu$ versus $T$ and $B$, respectively. Supplementary Information S12 shows that $z\nu$ diverges with decreasing temperature, indicating an enhanced effect of the quenched disorder, which introduces locally ordered superconducting rare regions on the microscopic level[5,14,15]. Fig. 3c displays $z\nu$ vs. $B$, which are found to follow the activated scaling law $z\nu = C|B_c^* - B|^{-\nu\psi}$. Here $C$ is a constant, $\nu\psi = 0.6$ is the 2D infinite-randomness critical exponent and $B_c^* = 7.9T$ is the divergent critical field, as obtained from the fitting. The value of $\nu\psi$ in 2D systems with infinite-randomness are predicted to be ~0.6 ($\nu \approx 1.2$ and $\psi \approx 0.5$)[33,34], which agrees well with our experiment, providing strong evidence for the existence of QGS in quasi-2D sample C. Additional data on QGS in other quasi-2D samples B and D are presented in Supplementary Information S5-8. We note that most of the Fe-based and Cu-based high-$T_c$ superconductors exhibit only one QCP in their SMT/SIT[2,8], with the only exception of underdoped $La_{2-x}Sr_xCuO_4$ films that exhibit two QCPs[4]. Thus, our data represents the first discovery of quasi-2D unconventional high-$T_c$ superconductors to host QGS in their superconductor-metal quantum phase transitions.

Despite the fact that QGS is rarely found in quasi-2D unconventional high-$T_c$ superconductors, QGS in 3D superconducting systems is even more elusive due to experimental and theoretical difficulties[41]. Fig. 3d shows MR of 3D anisotropic sample A versus $B_\perp$ at temperatures ranging from 0.49 K to 5.75 K; MR versus $B_{//}$ of sample A can be found in Supplementary Information S14. Surprisingly, the MR isotherms exhibit continuous movement of crossing points as sample temperature changes, indicative of QGS behavior of sample A under both $B_\perp$ and $B_{//}$. Fig. 3e and the inset of Supplementary Information S14b show $B_c$ versus $T$ for $B_\perp$ and $B_{//}$ configurations, respectively, which show similar deviation from the WHH theory at the low $T$ regime. FSS analysis is carried out for data collected from 3D anisotropic sample A (details in Supplementary Information S15-S16). Fig. 3f shows the resulting $z\nu$ vs. $B_\perp$ curve for sample A, following the same activated scaling law with $B_{c\perp}^* = 4.07$ T and $\nu\psi \approx 0.4$. Numerical simulations of 3D random quantum magnets[31,32] have predicted a correlation length exponent $\nu \approx 0.98$ and a tunneling critical exponent $\psi \approx 0.46$, resulting in $\nu\psi \approx 0.45$, in close agreement with our experiment. The analysis of $z\nu$ vs. $B_{//}$ for sample A can be found in Supplementary Information S14b. The critical behavior of sample A under $B_⊥$ and $B_{//}$ is found to have only two differences: 1) $B_c$ under $B_⊥$ diverges with decreasing temperature at the low-$T$ limit, while $B_c$ under $B_{//}$ appears to saturate at low temperatures. This phenomenon will be discussed in details in the Discussion Section. 2) the $B_c^*$ for the diverging $zv$ is 4.07 T for $B_⊥$ (Fig. 3f) and 12.41 T for $B_{//}$ (Supplementary Information S14b), highlighting the weakly anisotropic nature of the 3D electronic structure of sample A. Apart from these two differences, the behavior of sample A under $B_⊥$ and $B_{//}$ is essentially the same, including the fitted critical exponents $v\psi ≈ 0.4$, which are both in agreement with numerical simulations[31,32]. This remarkable consistency between the experimental data and the numerical simulations provides compelling evidence for the presence of QGS in the 3D unconventional high-$T_c$ superconductor CaFe$_{1-x}$Ni$_x$AsF ($x$ = 3.1%), representing the first experimental demonstration of magnetic field driven QGS of SMT in 3D systems.

***B-T* phase diagram and discussions**

Based on data from sample A, *B-T* phase diagrams of bulk 3D anisotropic unconventional Fe-based superconducting system with QGS under $B_⊥$ (Fig. 4a) and $B_{//}$ (Fig. 4b) are constructed. The phase diagrams are characterized by three curves: 1) the superconducting onset $T_c^{onset}$ $vs.$ $B$ curve (purple squares) which coincides with the $B_c$ $vs.$ $T$ curve (orange dots); 2) the mean field WHH[40] upper critical field $B_{c2}^{fit}$ $vs.$ $T$ curve (red dashed line); 3) the upper critical field $H_{c2}$ $vs.$ $T$ curve (blue dots, extracted form 90% of the normal resistance). These three curves divide the phase diagram into four regions: the SC state, the fluctuation region (including "QF": quantum fluctuations and "TF": thermal fluctuations), the QGS state and the normal state.

As shown in Fig.4a, the normal state in the diagram behaves as weakly localized metal above the overlapping points of $T_c^{onset}(B)$ and $B_c(T)$. The fluctuation region lies between the mean-field $B_{c2}^{fit}$ $vs.$ $T$ curve (red dashed line) and the $H_{c2}$ $vs.$ $T$ curve (blue dashed line). At low $T$, the $T_c^{onset}(B)$ curve turns upwards and significantly deviates from the mean-field WHH[40] behavior below a temperature $T_M$ ~ 1.9 K, indicating the emerging quantum fluctuation below $T_M$[21,42]. In particular, for temperatures below $T_M$ and for magnetic field above the mean-field WHH limit[40], the quantum Griffiths state emerges, characterized by a diverging $zv$ and an upturning $B_c(T)$ at low temperatures[5]. This quantum Griffiths state can be regarded as the effect related to quenched disorder on the Abrikosov vortex lattice in the region of $B_{c2}^{\text{fit}} < B < B_c^*$, where rare regions of large SC puddles appear when $T < T_M$. Meanwhile, the exponentially small excitation energy causes ultraslow dynamics accompanied by diverging effective dynamical exponent around zero $T$, similar to the behavior of large clusters in the random transverse field Ising model[43]. It is worth noting that QGS in samples A-D are conspicuously robust[18] (shown in Fig. 4c), with $T_M \sim 1.8$ K in sample A to $T_M \sim 5.3$ K in sample D, the later higher than any reported values in the literature (more details in Supplementary Information S7c), highlighting the peculiarity of QGS in unconventional high-$T_c$ superconductors.

Fig. 4b depicts the $B$-$T$ phase diagram of the same sample under $B_{//}$, which exhibits behavior similar to that under $B_\perp$, with one key difference: the $T_c^{\text{onset}}(B)$ or $B_c(T)$ under $B_{//}$ saturates at the low-$T$ limit, resulting in a narrow QGS region, compares to a diverging $T_c^{\text{onset}}(B)$ or $B_c(T)$ under $B_\perp$ with decreasing $T$. Saturating $B_c(T)$ has been previously reported in only three superconducting systems, the $B_{//}$-driven QGS of SMT in few-layer PdTe$_2$ films[23] and in $\beta$-W films[21], as well as the $B_\perp$-driven QGS of SIT in TiO films[19]. The former two cases[21,23] are considered to be QGS without the formation of vortex glass state, such that the low-$T$ divergent $B_c(T)$ is absent. For the third case, the saturated $B_c(T)$ is attributed to weaker Josephson coupling of the local rare regions in an insulating normal state background[19]. In our case, since the resistance of sample A is much less than previous reports of SMT/SIT in QGS[19,21,23], a saturating $B_c(T)$ cannot be explained by weaker Josephson coupling[19]; the anisotropic nature of the 3D electronic state in the material, on the other hand, might results in the $B_{//}$-induced rare regions without the emergence of the vortex glass state[23], leading to a saturating $B_c(T)$.

**Summary**

In summary, 3D quantum Griffiths singularities have been experimentally observed in the superconductor-metal transition of unconventional high-$T_c$ superconductor CaFe$_{1-x}$Ni$_x$AsF ($x$ = 3.1%) single crystals. A comprehensive quantum phase diagram for the 3D-QGS of SMT was established, which demonstrates the universality and similarity of QGS in 2D and 3D SC systems.

The robustness of QGS in FeAs-based SCs has also been confirmed. This work opens a new venue for exploring QGS physics in Fe/Cu-based high-$T_c$ superconductors and in 3D superconducting systems.

**Methods**

**Synthesis of $CaFe_{1-x}Ni_xAsF$ single crystals**

In this study, single crystals of $CaFe_{1-x}Ni_xAsF$ were synthesized using the CaAs self-flux method. Initially, a mixture of Ca granules and As grains in a 1:1 ratio was heated at 700°C for 10 hours in an evacuated quartz tube to obtain the CaAs precursor. The CaAs precursor was then subjected to a grinding and sintering process, repeated three times to ensure complete mixing and uniformity of CaAs, which is critical for obtaining high-quality single crystals. Subsequently, Fe powder, Ni powder, $FeF_2$ powder, and the homemade CaAs flux were mixed together in a stoichiometric ratio of 1-$x$: $x$: 1:15 ($x$ = 4%, 6%, 8%, 10%), placed in an alumina crucible, and sealed in a quartz tube under vacuum. The sealed quartz tube was then heated at 950°C and 1230°C for 45 hours and 30 hours, respectively, to promote crystal growth. Finally, the tube was cooled down to 850°C at a rate of 2°C/h, followed by quick cooling to room temperature.

**Characterization of $CaFe_{1-x}Ni_xAsF$ superconducting single crystals**

The $CaFe_{1-x}Ni_xAsF$ crystals exhibited a tetragonal structure, consisting of alternating CaF and $(Fe_{1-x}Ni_x)As$ layers along the $c$-axis, as depicted in Supplementary Information S1a. A typical high quality $CaFe_{1-x}Ni_xAsF$ single crystal (sample C, $x$ = 3.7%) has rectangular shape and approximately 4 x 3 x 0.15 mm in size, as shown in the inset of Supplementary Information S1b. X-ray diffraction of the $CaFe_{1-x}Ni_xAsF$ crystals (Supplementary Information S1c) confirmed the high crystallinity of the as-grown crystals with (001) orientation, consistent with the parent CaFeAsF compound[44]. Further characterization using scanning electron microscopy and energy dispersive spectrometer analysis (details in Supplementary Information S2) revealed flat as-grown (001) surfaces with definite Ni contents for different crystals. The above results prove the first successful growth of high-quality $CaFe_{1-x}Ni_xAsF$ single crystals.

**Transport Measurements**

Transport measurements were conducted in the temperature range of 0.3 K to 300 K, under magnetic fields of up to 14 T, using both an Oxford Teslatron cryostat and a Quantum Design PPMS. The resistance was measured at a frequency of 17.77 Hz using conventional lock-in techniques.

## Data Availability Statement

Source data are provided with this paper. Data for figures that support the current study are available at doi://xxx.

## Acknowledgements

This project has been supported by the National Key R&D Program of China (Grant No. 2019YFA0308402), the Innovation Program for Quantum Science and Technology (Grant No. 2021ZD0302403), the National Natural Science Foundation of China (NSFC Grant Nos. 11934001, 92265106, 11921005), Beijing Municipal Natural Science Foundation (Grant No. JQ20002). J.-H.C. thanks X.C.X. and H. L. for stimulating discussions. J.-H.C. acknowledges technical supports form Peking Nanofab.

## Author contributions

J.-H.C. & S.L. conceived the idea; J.-H.C. directed the experiment; S.L., Y.C. & C. C. provide high quality crystals; S.L. fabricated most of the devices and performed the transport measurements; C.T., H.C., X.W., M.C., J.L., Y.F., J.C., Y.Z., Y.S. & T.H. aided in sample preparation and transport measurement; J.-H.C., S.L., C.T., Y.Z. & Y.S. collected and analyzed the data; S.L. & J.-H.C. wrote the manuscript; all authors commented and modified the manuscript.

## Corresponding author

Correspondence to Jian-Hao Chen <chenjianhao@pku.edu.cn>

## Competing interests

The authors declare no competing interests.

After the submission of this work, we noticed reports of 3D-QGS in conventional three-dimensional $MgTi_2O_4$ superconducting film [arXiv:2311.06710] and MoTiN superconducting film [arXiv:2402.01347].

**Figures**

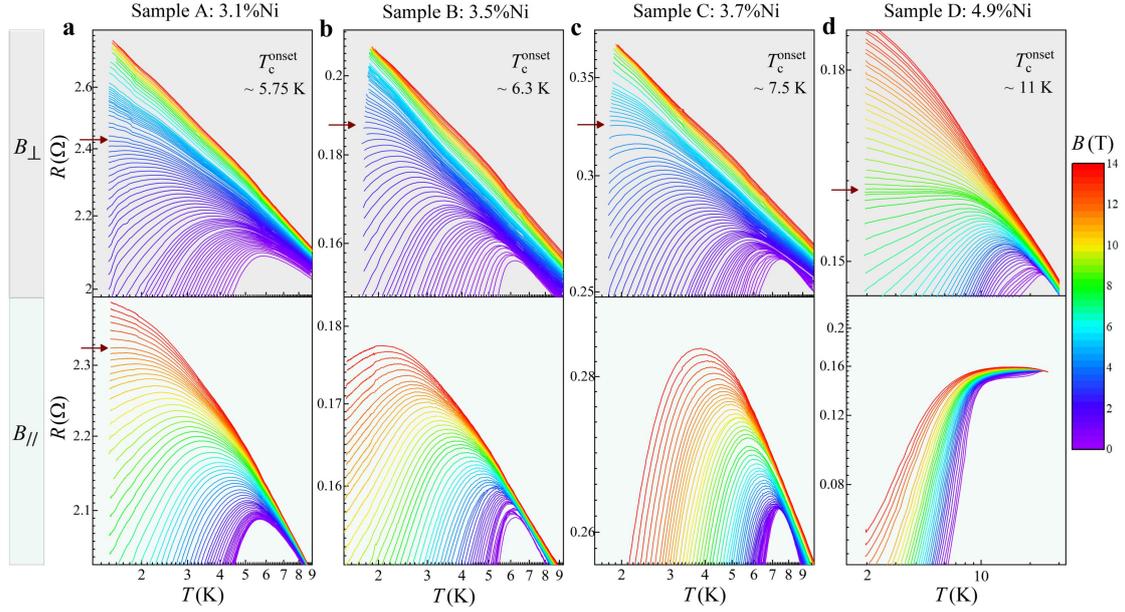

**Fig. 1. Magnetic-field-driven SMT with multiple QCPs in CaFe$_{1-x}$Ni$_x$AsF. a-d**, Temperature-dependent resistance under $B_\perp$ (upper panels) and $B_{//}$ (lower panels) from 0 to 14 T for CaFe$_{1-x}$Ni$_x$AsF samples A, B, C and D in log-log scale, respectively. Different colors represent different magnetic field values. The nickel (Ni) contents and the onset superconducting transition temperature for each sample are marked near the top of the figures. The brown arrows indicate the magnetic field where the field tunable SMTs occur.

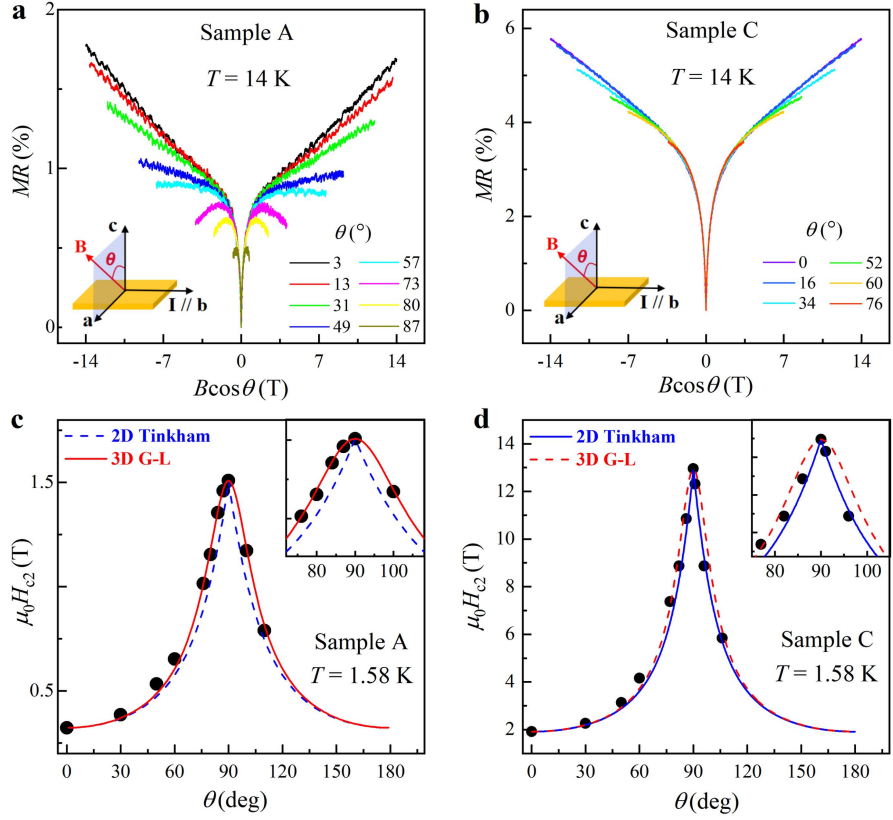

**Fig. 2. Doping tunable 3D anisotropic to quasi-2D crossover in CaFe$_{1-x}$Ni$_x$AsF single crystals. a,b,** Normal state MR vs. perpendicular magnetic field $B\cos\theta$ for sample A and C at 14 K, respectively. The magnetic field angle $\theta$, defined in the insets, have values 3°, 13°, 31°, 49°, 57°, 73°, 80° and 87° for sample A and 0°, 16°, 34°, 52°, 60° and 76° for sample C. **c,d,** Angular dependence of the upper critical fields $\mu_0 H_{c2}$ for sample A and C, respectively. The blue and red dashed lines are the theoretical curves of $H_{c2}(\theta)$ from the 2D Tinkham model[38] and the 3D anisotropic mass model[37], respectively. Insets show the zoom-in plot near $\theta = 90°$. $H_{c2}$ is defined as the magnetic field where the resistance becomes 50% of the normal state resistance.

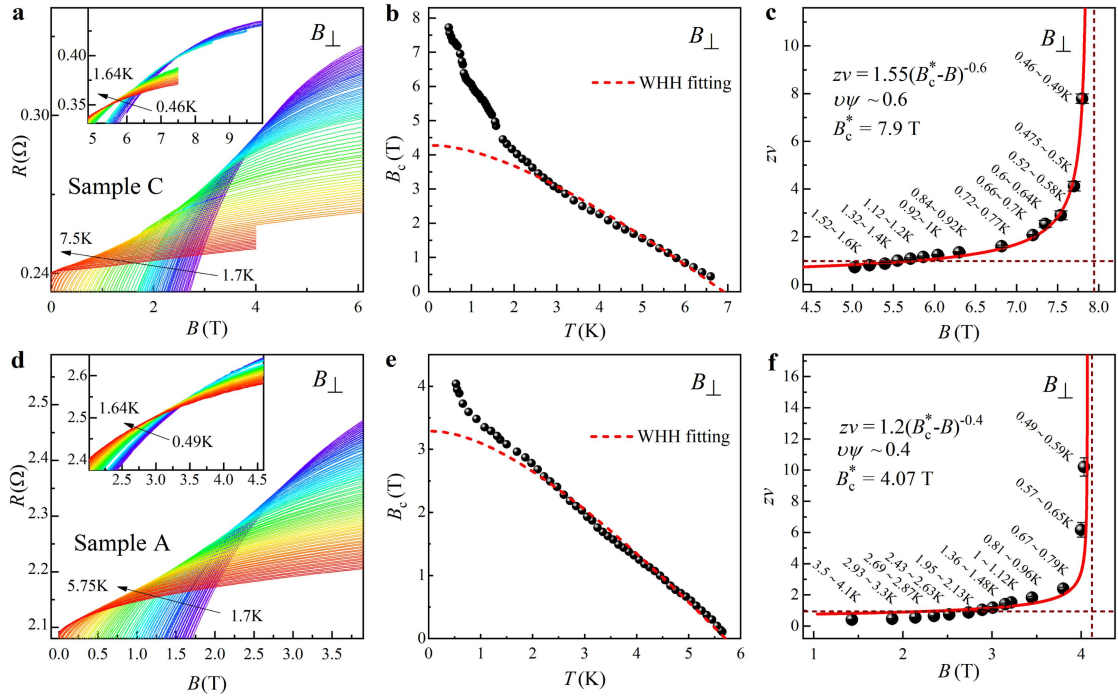

**Fig. 3. QGS of SMT in quasi-2D and 3D anisotropic CaFe$_{1-x}$Ni$_x$AsF single crystals. a**, $B_\perp$ dependent resistance isotherms for sample C at $T$ = 1.7–7.5 K in the main panel and 0.46–1.64 K in the inset, respectively. **b**, The critical magnetic field $B_c(T)$ extracted from **a**, the red dashed line is the fitting curve from the WHH theory[40]. **c**, The activated quantum scaling behavior of critical exponent $zv$ vs. $B_\perp$ in sample C. The red solid line is the fitting curve from the activated scaling law and gives $B_c^* = 7.9T$ (vertical dashed line). The horizontal dashed red line shows $zv = 1$. **d–f**, Similar data for sample A as in **a–c** for sample C under $B_\perp$.

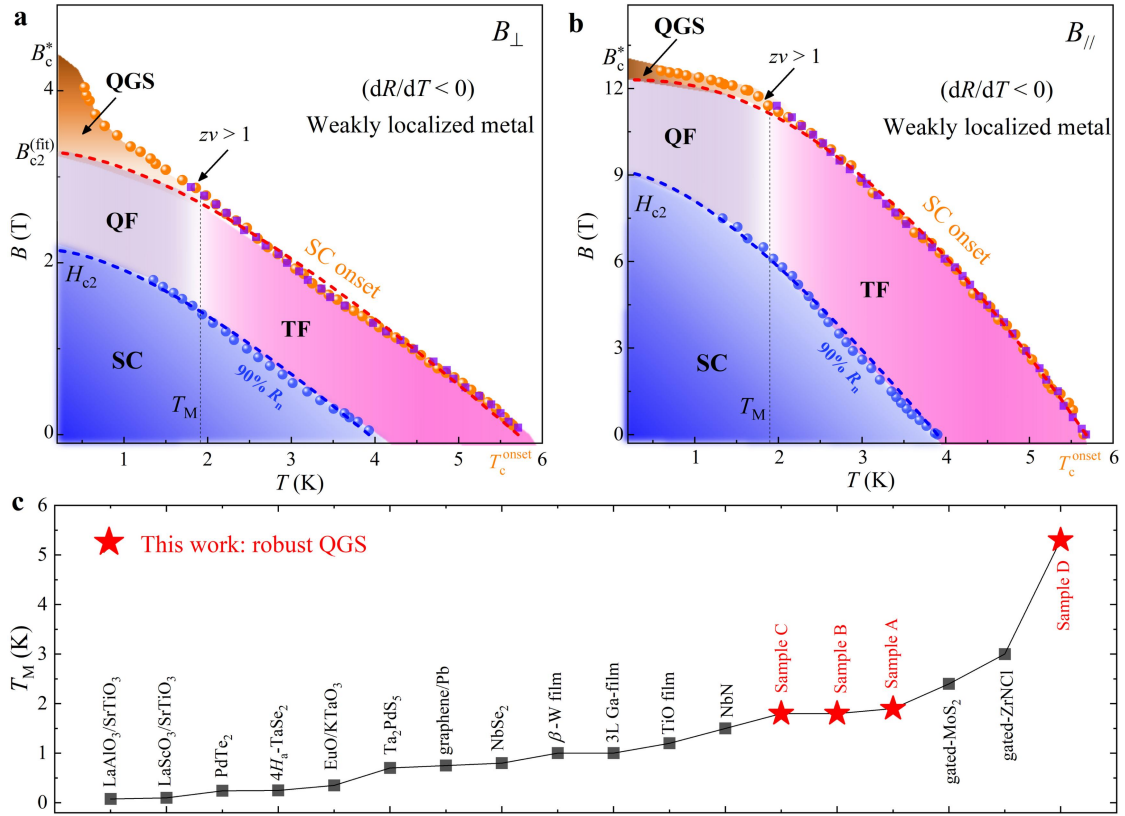

**Fig. 4.** *B–T* **phase diagram of QGS in a 3D anisotropic superconductor. a,b,** The phase diagrams are constructed with data from CaFe$_{1-x}$Ni$_x$AsF sample A under $B_\perp$ and $B_{//}$, respectively. The QGS phase is the new discovery from this work. Purple squares and orange dots: $T_c^{onset}(B)$ from $R(B)$ vs. $T$ curves and $B_c$ vs. $T$ curves, respectively. Blue dots: $H_{c2}$ vs. $T$ defined as 90% of the normal state resistance. Red and blue dashed lines: fitted curve of $T_c^{onset}(B)$ to the WHH theory[40]. "QF" and "TF" denotes quantum fluctuations and thermal fluctuations, respectively. **c,** Robust QGS in CaFe$_{1-x}$Ni$_x$AsF single crystals. We plotted the QGS emerging temperature $T_M$ for sample A, B, C and D together with $T_M$ for all the other reported QGS superconductor-metal transitions in the literature [refs.[5,7,17-21,23,45-49]]. It can be seen that $T_M$ for CaFe$_{1-x}$Ni$_x$AsF single crystals are much higher than other reported values, underscoring the robustness of the QGS in unconventional high-$T_c$ superconductors.